\documentclass[aps,prd,onecolumn,groupedaddress,showpacs,nofootinbib,amssymb]{revtex4}
\usepackage[dvips]{graphicx}
\usepackage{amssymb}
\usepackage{amsmath}
\usepackage{graphicx,,color}
\usepackage{amsfonts}
\usepackage{bm}
\usepackage{cancel}
\usepackage{comment}

\newcommand\be{\begin{equation}}
\newcommand\ee{\end{equation}}

\allowdisplaybreaks[4]

\begin{document}

\title{Amplification of Primordial Gravitational Waves by a Geometrically Driven non-canonical Reheating
Era}
\author{S.D. Odintsov,$^{1,2}$}
\email{odintsov@ice.cat}\author{V.K.
Oikonomou,$^{3,4}$}\email{v.k.oikonomou1979@gmail.com,voikonomou@auth.gr}
\affiliation{$^{1)}$ ICREA, Passeig Luis Companys, 23, 08010 Barcelona, Spain\\
$^{2)}$ Institute of Space Sciences (IEEC-CSIC) C. Can Magrans
s/n,
08193 Barcelona, Spain\\
$^{3)}$Department of Physics, Aristotle University of Thessaloniki, Thessaloniki 54124, Greece\\
$^{4)}$ Laboratory for Theoretical Cosmology, Tomsk State
University of Control Systems and Radioelectronics  (TUSUR),
634050 Tomsk, Russia}

\tolerance=5000

\begin{abstract}
In order to describe inflation in general relativity, scalar
fields must inevitably be used, with all the setbacks of that
description. On the other hand, $f(R)$ gravity and other modified
gravity theories seem to provide a unified description of early
and late-time dynamics without resorting to scalar or phantom
theories. The question is, can modified gravity affect directly
the mysterious radiation domination era? Addressing this question
is the focus in this work, and we shall consider the case for
which in the early stages of the radiation domination era, namely
during the reheating era, the background equation of state
parameter is different from $w=1/3$. As we show, in the context of
$f(R)$ gravity, an abnormal reheating era can affect the
primordial gravitational wave energy spectrum today. Since future
interferometers will exactly probe this era, which consists of
subhorizon modes that reentered the horizon during the early
stages of the radiation domination era, the focus in this work is
how a short abnormal reheating era that deviates from the standard
perfect fluid pattern with $w\neq 1/3$, and generated by higher
order curvature terms, can affect the primordial gravitational
wave energy spectrum. Using a WKB approach, we calculate the
effect of an $f(R)$ gravity generated abnormal reheating era, and
as we show the primordial gravitational wave spectrum is
significantly amplified, a result which is in contrast to the
general relativistic case, where the effect is minor.
\end{abstract}


\maketitle

\section{Introduction}

To date general relativity (GR) seems to be a successful
description of the Universe at astrophysical levels, and only a
few events seem to deviate from the standard GR description at an
astrophysical levels \cite{Abbott:2020khf}, still this deviation
is debatable. It however seems to be difficult for GR to
consistently describe large scales of the Universe at a
cosmological level, since dark energy cannot be described in a
rigid and non-problematic way by GR. Specifically, a dark energy
era with phantom divide crossing in GR would require the presence
of phantom scalar fields, a feature not so desirable in
theoretical physics models. On the other hand, the inflationary
era \cite{inflation1,inflation2,inflation3,inflation4} can be
described in the context of GR, but again the presence of a scalar
field which drives the evolution is required. Although scalar
fields are in general expected to be present in the low-energy
regime of some fundamental underlying string theory, these are the
string moduli, the description of inflation with some fundamental
scalar other than the Higgs field has some shortcomings. Indeed,
the inflaton particle properties, like its mass, its coupling  to
fundamental particles, its potential energy, make difficult the
identification of the inflaton in terms of fundamental particle
physics models. In some sense the inflaton is introduced in an ad
hoc way, and its properties seem to be the result of a suitable
fine tuning of its parameters relevant to inflation.

A natural extension of GR which overcomes the scalar field
problems we mentioned, is offered by modified gravity in its
various forms
\cite{reviews1,reviews2,reviews3,reviews4,reviews5,reviews6}. The
extension of GR is a natural choice, since the fundamental
curvature term is present in the Einstein-Hilbert Lagrangian, thus
it is possible that higher order curvature terms might be present,
originating from the underlying fundamental quantum gravity
theory. With modified gravity it is possible to describe both the
inflationary era and the dark energy era, without the introduction
of scalar fields, see for example
\cite{Nojiri:2003ft,Capozziello:2005ku,Hwang:2001pu,Cognola:2005de,Song:2006ej,Faulkner:2006ub,Olmo:2006eh,Sawicki:2007tf,Faraoni:2007yn,Carloni:2007yv,
Nojiri:2007as,Deruelle:2007pt,Appleby:2008tv,Dunsby:2010wg} for
works in the context of the most fundamental modified gravity
theory, namely $f(R)$ gravity.

While in most cases the focus in modified gravity theories is on
inflation and dark energy and how these two eras can be realized
by modified gravity, the question is how does modified gravity
affect the intermediate eras, namely the radiation and the matter
domination era. With regard to the latter, post-recombination we
know that the Universe should mimic the $\Lambda$-Cold-Dark-Matter
($\Lambda$CDM) model, however before that and specifically during
the radiation domination era, the plot thickens. In principle, the
radiation domination era, which starts with the reheating era, is
quite mysterious. We have no clue on what actually happened in the
Universe during this evolutionary patch of it. We have theoretical
hints and theoretical requirements, but this era is quite
mysterious and the temperature is quite large of the order
$T>150$GeV. Hence the theoretical question is, can higher
curvature correction terms affect in some way this era? Can
modified gravity have a direct imprint on some stages of this
evolutionary era? Usually it is assumed that $w=1/3$ during
radiation era, which is a not however a necessary assumption, just
a theoretical expectation. In fact, for this era, the background
EoS parameter $w$ can be found in the range $0<w<1/3$
\cite{Boyle:2005se}. Hence, an abnormal reheating era with $w\neq
1/3$ can have implications on the present day primordial
gravitational waves energy spectrum, if the relevant modes
reentered the horizon during this abnormal reheating era. In the
context of GR however the modification of the gravitational wave
energy spectrum is minor, so the question is whether this pattern
is repeated in modified gravity. Specifically, if a modified
gravity generates a short abnormal reheating era, is the
gravitational wave energy spectrum significantly affected? The
answer lies in the affirmative for $f(R)$ gravity as we will show
in this paper. The motivation for modifying the reheating era is
two fold: Firstly it is theoretically more likely for the higher
order curvature terms to operate during this mysterious era, and
secondly, it is the era where the modes that will be probed by
future high frequency experiments reenter the horizon after
inflation. Thus if an abnormal reheating era occurs, this will be
detectable and perhaps verified.

To be specific, in order to discriminate which model drives a
possible future observation, several experiments must be combined.
Since inflation plays a prominent role in future experiments, it
will be tested by both Cosmic Microwave Background (CMB)
experiments \cite{CMB-S4:2016ple,SimonsObservatory:2019qwx} and by
high frequency interferometers
\cite{Hild:2010id,Baker:2019nia,Smith:2019wny,Crowder:2005nr,Smith:2016jqs,Seto:2001qf,Kawamura:2020pcg,Bull:2018lat}.
These modes will verify either the existence of a $B$-mode pattern
in the CMB temperature fluctuations \cite{Kamionkowski:2015yta},
or they will verify the existence of a stochastic primordial
gravitational wave background. The CMB experiments will hopefully
measure the tensor-to-scalar ratio and possibly will determine the
tensor spectral index. If the tensor spectral index is red-tilted
and at the same time a stochastic cosmological signal is observed
by the gravitational waves experiments, the theories that can
describe such a physical situation are quite narrowed down. In
fact, single scalar field theories and their Jordan frame
counterpart theories, namely $f(R)$ gravities, will be too
difficult to describe the physics, if not impossible, assuming a
standard reheating era. However, with the present paper we want to
stress the fact that if an abnormal reheating era is generated by
$f(R)$ gravity, the amplification of the primordial gravitational
wave energy spectrum is significant enough to be detected by most
of the future experiments, even if the abnormal reheating era
lasts for a short period of time after the end of inflation.
Hence, the plot seems to thicken in modified gravity cosmology, if
such alternative reheating-radiation domination era scenarios are
taken into account. In this work we shall describe the different
evolutionary patches of the Universe using an underlying $f(R)$
gravity. The inflationary patch, if it is a quasi-de Sitter
evolution, then it will be generated by an $R^2$ gravity, and
matter and radiation fluids are ignored. We shall assume that
during early stages of the radiation domination era, specifically
during the reheating era, a short period of abnormal reheating
takes place, with the total EoS parameter being $w\neq 1/3$, in
the presence of radiation and dark matter perfect fluids. We find
which $f(R)$ gravity can generate such an evolution and
accordingly we examine the matter and late-time eras of the
Universe, which are controlled by an appropriate $f(R)$ gravity
term, so as for the $f(R)$ model to mimic the $\Lambda$CDM model.
This late-time $f(R)$ gravity term does not affect at all the
inflationary nor the abnormal reheating era. Accordingly, we find
numerically the amplification of the primordial gravitational wave
energy spectrum caused by this abnormal reheating era, and as we
show it is quite large to make the signal detectable by all future
experiments. In fact, the larger the duration of the abnormal
reheating is, the larger the amplification of the signal becomes.

This paper is organized as follows: In section II we discuss the
various evolutionary patches of the Universe, from inflation to
dark energy and discuss how these can be realized by $f(R)$
gravity in the presence of dark matter and radiation fluids.
During inflation, the matter fluids are neglected, but
post-inflationary these must be taken into account. We also add a
short period of abnormal reheating with EoS parameter $w=0.1$ and
we examine which $f(R)$ gravity in the presence of dark matter and
radiation perfect fluids can realize such an exotic epoch. The
late-time properties of the proposed $f(R)$ gravity is examined
too. In section III we calculate the effects of the abnormal
reheating era on the primordial gravitational wave energy
spectrum. A discussion along with the conclusions follow in the
conclusions section.

\section{Evolution of the Universe with $f(R)$ Gravity: From Inflation to the Dark Energy Era and non-canonical Reheating Realization}

To date, the basic assumptions for the evolution of the Universe
mainly concern four evolutionary regimes, the inflationary era,
the reheating-radiation domination era, the matter domination era
and the late-time acceleration era, the so-called dark energy era.
The question how the Universe smoothly passes through these era is
not firmly answered, although several appealing models can provide
a unified description of most of the Universe's evolution eras.
Modified gravity in its various forms seems to be an inevitable
choice for describing the inflationary and dark energy era, mainly
the latter to be honest, since GR fails to describe consistently
the late-time era. Also in the context of GR, the inflationary era
description relies on scalar fields, and this has several
shortcomings as we mentioned in the introduction. Thus, modified
gravity seems to provide a consistent description for the
inflationary era and the dark energy era, the two acceleration
eras of the Universe. But the question then is, which other eras
may be described by modified gravity, or equivalently, why should
modified gravity affect only inflation and the dark energy eras?
In principle it should be present and control all the evolutionary
eras. For the matter domination era, it should provide a
$\Lambda$CDM like evolution for small redshifts near the end of
the matter domination eras, and the major question then is what it
happens during reheating-radiation domination era. The reheating
and the evolution of it to the radiation domination era, is quite
mysterious, and we know almost nothing for this era. The only way
to have our grasp on it will be offered by future interferometer
experiments like the LISA mission, or DECIGO, which will probe
frequencies corresponding to inflationary modes that reentered the
horizon during the reheating-radiation domination eras. If an
underlying $f(R)$ gravity controls the evolution of the Universe,
synergistically with cold dark matter and radiation fluids
post-inflationary, then the $f(R)$ gravity should have its
imprints during both the radiation and the matter domination eras.
Among the matter and radiation domination eras, it is highly
likely that the total EoS of the Universe might be non-standard
during the early stages of the radiation domination era, hence
during reheating and beyond. During the matter domination era, the
effects of the underlying $f(R)$ gravity should provide a
$\Lambda$CDM model like evolution, at least for small redshifts
post-recombination.

Hence, if exotic scenarios should occur, it is highly likely that
these occurred post-inflationary and during the early stages of
the radiation domination era. This is the scenario that we will
describe in this work, in the context of $f(R)$ gravity. Although
during inflation, the effects of cold dark matter and radiation
fluids are neglected, post-inflationary these fluids cannot be
neglected. Hence our approach for the post-inflationary early
stages of the reheating era, should include radiation and cold
dark matter fluids. Finding the exact form of the underlying
$f(R)$ gravity which describes all the evolution eras of our
Universe, is a rather formidable task, however we can find the
leading order term of $f(R)$ gravity which can describe the
different patches of the Universe's evolution. During the
inflationary era, the Universe is described by a quasi-de Sitter
evolution for example, which by neglecting the matter fields, as
we show can be described by an $R^2$ gravity. Accordingly, for the
radiation domination era we shall assume that the Universe has a
constant EoS parameter $w$ different from that of radiation.

Let us start with the gravitational action of $f(R)$ gravity in
the presence of perfect matter fluids,
\begin{equation}\label{action1dse}
\mathcal{S}=\frac{1}{2\kappa^2}\int
\mathrm{d}^4x\sqrt{-g}f(R)+\mathcal{S}_m,
\end{equation}
with $\kappa^2$ denoting as usual $\kappa^2=8\pi
G=\frac{1}{M_p^2}$, where $G$ is Newton's gravitational constant
and $M_p$ stands for the reduced Planck mass. In the metric
formalism, the field equations can be found by varying the action
with respect to the metric, and these are,

\begin{equation}\label{eqnmotion}
f_R(R)R_{\mu \nu}(g)-\frac{1}{2}f(R)g_{\mu
\nu}-\nabla_{\mu}\nabla_{\nu}f_R(R)+g_{\mu \nu}\square
f_R(R)=0\kappa^2T_{\mu \nu}^{m}\, ,
\end{equation}
where $T_{\mu \nu}^m$ is the energy momentum tensor of the matter
perfect fluids, and we introduced
$f_R=\frac{\mathrm{d}f}{\mathrm{d}R}$. For a flat
Friedmann-Robertson-Walker (FRW) metric, the Friedmann equation
becomes,
\begin{equation}\label{frwf1}
-18\left (4H(t)^2\dot{H}(t)+H(t)\ddot{H}(t)\right)f_{RR}(R)+3
\left(H^2(t)+\dot{H}(t)
\right)f_(R)-\frac{f(R)}{2}+\kappa^2\left(\rho_m+\rho_r
\right)=0\, ,
\end{equation}
where $\rho_m$, $\rho_r$ denote the energy density of the cold
dark matter and radiation respectively.

Let us now consider the primordial patch of the Universe's
evolution, which comprises from the inflationary patch and the
radiation domination era. We shall assume that the early stages of
the radiation domination era is described by a constant EoS $w$,
so the evolution during the early radiation era is basically a
power-law. Also during the inflationary patch, the Universe is
described by a quasi-de Sitter evolution, so the scale factor
during the inflationary and early post-inflationary era is the
following,
\begin{equation}\label{scalefactorquasidesitter}
a(t)=a_0e^{H_0 t-H_i^2 t^2}+a_{r}t^{\frac{2}{3(1+w)}}\, ,
\end{equation}
where $a_0$ is the scale factor at the beginning of the
inflationary era, and $a_r$ is the scale factor at the beginning
of the radiation era, and at the end of inflation, and $H_i$ are
free parameters with mass dimensions $[H_0]=[H_i]=[m]$. With
regard to the value of the total EoS parameter $w$, we shall
assume that it is not $w=1/3$, but it is similar to a deformed
matter domination era EoS parameter, so $w=0.1$. The whole
analysis works with other values different than $w=1/3$, but let
us fix $w=0.1$, so that the Universe does not commence the
radiation domination era with a pure $w=1/3$ but more closely to
the matter domination total EoS parameter $w=0$. In the
literature, the values of the total EoS parameter during the
reheating process, which is the early radiation domination era
period, are expected to be found in the range $0<w<\frac{1}{3}$
\cite{Boyle:2005se}. In fact, the value of the total EoS parameter
we chose, which is very close to the $w=0$ case, is considered
motivated in the literature \cite{Boyle:2005se} and in the context
of GR the effect of such EoS parameter on the primordial
gravitational wave energy spectrum would be similar to having
massive relics.

\begin{figure}[h!]
\centering
\includegraphics[width=16pc]{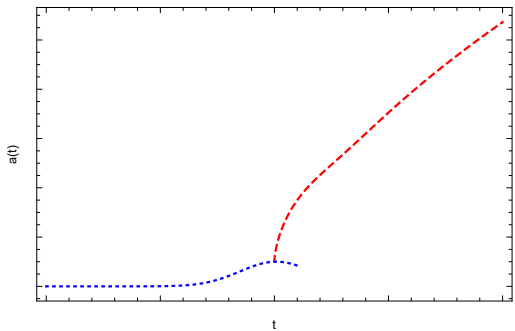}
\caption{The scale factor of the scale factor
(\ref{scalefactorquasidesitter}) (red curve) and for the pure
quasi de-Sitter evolution $a(t)\sim e^{H_0 t-H_i^2 t^2}$ (blue
curve).} \label{plot1}
\end{figure}

Coming back to the evolution (\ref{scalefactorquasidesitter}), it
describes a quasi de-Sitter evolution at early times, which is the
exponential part, followed by the power-law evolution which
describes the Universe during the early post-inflationary era. The
quasi de-Sitter evolution dominates at early times, during
inflation, and after that, the power-law evolution dominates. This
can be seen clearly in Fig. \ref{plot1} where with blue dotted
curve we plot the scale factor corresponding to the pure quasi-de
Sitter scalae factor $a(t)\sim e^{H_0 t-H_i^2 t^2}$, and with
red-dashed curve the scale factor of Eq.
(\ref{scalefactorquasidesitter}), where it is apparent that the
power-law part of the scale factor dominates after the quasi-de
Sitter era ends.
Finding the combined form of the $f(R)$ gravity that realizes the
combined evolution (\ref{scalefactorquasidesitter}) is a rather
formidable task, however we can find which $f(R)$ gravity realizes
the different patches of the evolution. So we can find the
dominant form of the $f(R)$ gravity which approximately generates
the quasi-de Sitter patch, and then we can find the approximate
$f(R)$ gravity which generates the power-law part of the scale
factor (\ref{scalefactorquasidesitter}). For finding the
approximate forms of the $f(R)$ gravity, we shall employ well
known reconstruction techniques which were developed in Ref.
\cite{Nojiri:2009kx}. The method is based on using the
$e$-foldings number as a dynamical variable instead of the cosmic
time, with the former being defined as follows,
\begin{equation}\label{efoldpoar}
e^{-N}=\frac{a_i}{a}\, ,
\end{equation}
where $a_i$ is defined as some initial value of the scale factor.
By using the $e$-foldings number $N$, the Friedmann equation takes
the form,
\begin{eqnarray}
\label{newfrw1} &&
\!\!\!\!\!\!\!\!\!\!\!\!\!\!\!\!\!\!\!\!\!\!\!\!\! -18\left [
4H^3(N)H'(N)+H^2(N)(H')^2+H^3(N)H''(N) \right ]f_{RR}(R)\notag
\\
&& \ \ \ \ \ \ \ \ \ +3\left [H^2(N)+H(N)H'(N)
\right]f_R(R)-\frac{f(R)}{2}+\kappa^2\rho=0\, ,
\end{eqnarray}
where $\rho=\rho_m+\rho_r$. Introducing the function
$G(N)=H^2(N)$, the Ricci scalar is written as,
\begin{equation}\label{riccinrelat}
R=3G'(N)+12G(N)\, .
\end{equation}
Now by specifying the scale factor and correspondingly the Hubble
rate, using Eq. (\ref{riccinrelat}) one can invert it and find the
function $N(R)$, thus upon substituting the resulting expression
in Eq. (\ref{newfrw1}), one ends up with a second order
differential equation with the dynamical variable being the Ricci
scalar $R$, which in the absence of perfect matter fluids, in
terms of $G(N)$ takes the form,
 \begin{equation}
\label{newfrw1modfrom} -9G(N(R))\left[ 4G'(N(R))+G''(N(R))
\right]f_{RR}(R) +\left[3G(N)+\frac{3}{2}G'(N(R))
\right]f_R(R)-\frac{f(R)}{2}=0\, ,
\end{equation}
where $G'(N)=\mathrm{d}G(N)/\mathrm{d}N$ and
$G''(N)=\mathrm{d}^2G(N)/\mathrm{d}N^2$ and since we consider the
quasi. Thus upon solving it, one may obtain the $f(R)$ gravity
which realizes the given scale factor.

Let us first consider the quasi de-Sitter part of the scale factor
(\ref{scalefactorquasidesitter}) $a(t)\sim e^{H_0-H_i^2 t^2}$, so
in this case we have,
\begin{equation}\label{gnfunction}
G(N)=H_0^2-4 H_i^2 N\, .
\end{equation}
Using Eqs. (\ref{riccinrelat}) and (\ref{gnfunction}), the
$e$-foldings number $N$ in terms of the Ricci scalar is found to
be,
\begin{equation}\label{efoldr}
N=\frac{12 H_0^2-12 H_i^2-R}{48 H_i^2}\, .
\end{equation}
Using the above, the Friedmann equation becomes,
\begin{align}
\label{bigdiffgeneral1} & \left(12 H_i^2 \left(12
H_i^2+R\right)\right)\frac{\mathrm{d}^2f(R)}{\mathrm{d}R^2} +
\left(\frac{R}{4}-3
H_i^2\right)R\frac{\mathrm{d}f(R)}{\mathrm{d}R}-\frac{f(R)}{2}=0,
\end{align}
the solution of which is,
\begin{equation}\label{frformprev}
f(R)=R+\frac{R^2}{72 H_i^2}+2 H_i^2-\frac{\mathcal{C}_2 \left(144
H_i^4+72 H_i^2 R+R^2\right) \left(\sqrt[4]{e} \sqrt{3 \pi }
\text{erf}\left(\frac{\sqrt{4 H_i^2+\frac{R}{3}}}{4
H_i}\right)+\frac{12 H_i e^{-\frac{R}{48 H_i^2}} \left(36
H_i^2+R\right) \sqrt{12 H_i^2+R}}{144 H_i^4+72 H_i^2
R+R^2}\right)}{3981312 H_i^9} \, ,
\end{equation}
where $\mathcal{C}_2$ has mass dimensions $[\mathcal{C}_2]=[m]^7$
and it is simply an integration constant. Apparently the model
(\ref{frformprev}) is a deformation of the $R^2$ model and it can
be shown (the calculation will be presented elsewhere), the model
is quantitatively a deformation of the $R^2$ model. Indeed, for
$N\sim 60$, and irrespective of the values of the free parameters,
the spectral index of the primordial scalar curvature
perturbations for this model is $n_s\sim 0.967078$, while the
tensor-to-scalar ratio and the tensor spectral index are
$r=0.00327846$ and $n_T\simeq -0.000135483$. All the values are
very close to the standard $R^2$ model. A term which can be added
in the inflationary and has an insignificant effect during
inflation, is the following,
\begin{equation}\label{frde}
f_{DE}(R)=\delta R e^{-\frac{\Lambda_1}{R}}e^{-15\Lambda_2 R}\, ,
\end{equation}
where $\delta$ is a dimensionless parameter, and $\Lambda_1$,
$\Lambda_2$ are parameters with cosmological constant dimensions
and will be chosen to be of the same order of magnitude. The term
(\ref{frde}) has an insignificant effect during inflation, thus
the quasi-de Sitter evolution is still generated by the $f(R)$
(\ref{frformprev}) effectively. However its effect at late times
is important, since as we show later on in this work, it can
generate a deformed $\Lambda$CDM evolution, while its contribution
during the radiation domination era is also insignificant.

Now we shall derive which $f(R)$ gravity can realize the power-law
evolutionary patch of the evolution
(\ref{scalefactorquasidesitter}). In this case we shall take into
account the presence of the cold dark matter and radiation, thus
the Friedmann equation has the form (\ref{newfrw1}). In the case
at hand, with the scale factor being of the form $a(t)\sim
t^{2/(3(w+1))}$, the function $G(N)$ reads,
\begin{equation}\label{gnfunction}
G(N)=\frac{4 e^{-3 N (w+1)}}{9 (w+1)^2}\, ,
\end{equation}
where we have set $a_r=1$ for convenience. Thus upon combining
Eqs. (\ref{riccinrelat}) and (\ref{gnfunction}), we can obtain the
$e$-foldings number $N$ as a function of $R$,
\begin{equation}\label{efoldr}
N=\frac{\log \left(\frac{4 (1-3 w)}{3 R (w+1)^2}\right)}{3 (w+1)}.
\end{equation}
  Inserting the above in the Friedmann equation  (\ref{newfrw1}),
  we can solve the differential equation, however before that
  there is an important step related with the perfect matter
  fluids. The total matter energy density $\rho_{tot}$ must be expressed in terms of
$N(R)$, thus it will be a function of the Ricci scalar eventually.
Since the matter and radiation fluids are perfect fluids, then the
energy densities $\rho_i$, $i=r,m$, satisfy independently the
continuity equation $\dot{\rho}_i+3H(1+w_i)\rho_i=0$, with $w_i$
their corresponding EoS parameters. Thus,
\begin{equation}\label{mattenrgydens}
\rho_{tot} =\sum_i\rho_{i0}a_0^{-3(1+w_i)}e^{-3N(R)(1+w_i)}\, ,
\end{equation}
and therefore, by inserting all the above in the Friedmann
equation (\ref{newfrw1modfrom}), we obtain the following
differential equation for $f(R)$,
\begin{align}
\label{bigdiffgeneral1} &a_1
R^2\frac{\mathrm{d}^2f(R)}{\mathrm{d}R^2}
+a_2R\frac{\mathrm{d}f(R)}{\mathrm{d}R}-\frac{f(R)}{2}+\sum_iS_{i}R^{
\frac{3(1+w_i) }{3(1+w)}}=0\, ,
\end{align}
where $a_1$ and $a_2$ are defined as follows,
\begin{eqnarray}
\label{apara1a2}
&&a_1=\frac{3(1+w)}{4-3(1+w)}\nonumber\\
&&a_2=\frac{2-3(1+w)}{2(4-3(1+w))},
\end{eqnarray}
and recall that the index ``i'' takes the values $i=(r,m)$ with
$i=r$ denoting radiation and $i=m$ the cold dark matter perfect
fluid. Also the parameters $S_i$ stand for,
\begin{equation}\label{gfdgfdgf}
S_i=\frac{\kappa^2\rho_{i0}a_0^{-3(1+w_i)}}{[3A(4-3(1+w))]^{\frac{3(1+w_i)}{3(1+w)}}}
\, ,
\end{equation}
and the parameter $A$ is defined as follows,
\begin{equation}\label{alhpaaux}
A=\frac{4}{3(w+1)}\, .
\end{equation}
The solution of the differential equation (\ref{bigdiffgeneral1})
provides the exact $f(R)$ gravity that produces the power-law
patch of the evolution (\ref{scalefactorquasidesitter}). The
general solution of the differential equation
(\ref{bigdiffgeneral1}) is easily found to be,
\begin{equation}
\label{newsolutionsnoneulerssss} f_{p}(R)=\left
[\frac{c_2\rho_1}{\rho_2}-\frac{c_1\rho_1}{\rho_2(\rho_2-\rho_1+1)}\right]R^{\rho_2+1}
+\sum_i
\left[\frac{c_1S_i}{\rho_2(\delta_i+2+\rho_2-\rho_1)}\right]
R^{\delta_i+2+\rho_2}-\sum_iB_ic_2R^{\delta_i+\rho_2}+c_1R^{\rho_1}+c_2R^{\rho_2}\,
,
\end{equation}
where $c_1,c_2$ arbitrary integration constants, and we defined
$\delta_i$ and $B_i$, $i=(r,m)$ as follows,
\begin{equation}
\label{paramefgdd}
\delta_i=\frac{3(1+w_i)-23(1+w)}{3(1+w)}-\rho_2+2,\,\,\,B_i=\frac{S_i}{\rho_2\delta_i}
\, .
\end{equation}
Thus the $f_{p}(R)$ gravity of Eq.
(\ref{newsolutionsnoneulerssss}) can generate an early radiation
era which is different from the GR pattern in which $w=1/3$. This
geometrically generated exotic early radiation domination era with
$w=0.1$ as we said, can last sufficiently long, but we shall
discuss this important issue in the next section. As we will show
in the next section, the geometrically generated exotic radiation
domination era, can affect significantly the energy spectrum of
the primordial gravitational waves, leading to an amplification of
the predicted signal at present day. An observation we made is
that the inclusion of the early dark energy term (\ref{frde}) does
not affect at all the resulting energy spectrum. Thus we can
safely assume that this early dark energy term does not affect the
evolution aspects of the model until late times, where it is
responsible for a viable $\Lambda$CDM-like dark energy era. Before
going to the late-time era analysis, let us summarize our
findings, and make some assumptions on the issue of the duration
of the exotic patch of the radiation domination era. Primordially
in our scenario, the Universe is described by a quasi-de Sitter
patch which is realized by an $R^2$ gravity, which is followed by
a short period of a non-canonical reheating, with the background
EoS parameter being $w$ instead of the standard $w=1/3$. In this
case, the dominant form of the $f(R)$ gravity which realizes this
short non-canonical reheating era is given by Eq.
(\ref{newsolutionsnoneulerssss}). It is conceivable that this
short abnormal reheating era is geometrically driven by $f(R)$
gravity and the physics of it is different compared to GR. This
feature will be strongly justified in the next section, where a
comparison to the GR pattern of effects shall be given. Also it is
notable that the addition of the dark energy term (\ref{frde})
does not affect at all both the inflationary era and the abnormal
short radiation domination era period. For the short abnormal
reheating era, this feature will be justified in the next section.
After the geometrically driven non-canonical reheating era, the
Universe enters the standard radiation domination era, followed by
the dark matter domination era, for which the dark energy $f(R)$
gravity term also affects the evolution, especially at late times.
We can give a schematic description of the different evolution
patches of our Universe, and the corresponding $f(R)$ gravity
description, which we quote below,
\begin{align}\label{fralltimes}
& T\sim 10^{16}\mathrm{GeV}-10^{12}\mathrm{GeV},\,\,\,f(R)\sim R+R^2+f_{DE}(R)\, ,\\
\notag &
T\sim 10^{12}\mathrm{GeV}-10^{10}\mathrm{GeV},\,\,\,f(R)\sim f_{p}(R)+f_{DE}(R)+\mathrm{radiation}\,\,\mathrm{and}\,\,\mathrm{dark}\,\,\mathrm{matter}\,\,\mathrm{fluids}\, ,\\
\notag & T<10^{10}\mathrm{GeV},\,\,\,f(R)\sim
R+f_{DE}(R)+\mathrm{radiation}\,\,\mathrm{and}\,\,\mathrm{dark}\,\,\mathrm{matter}\,\,\mathrm{fluids}\,
,\, .
\end{align}
In the schematic description (\ref{fralltimes}) we assumed that
the short abnormal reheating era lasts for a short period of time,
in the temperature range $10^{12}$GeV to $10^{10}$GeV. So this
geometrically driven era occurs at the beginning of the radiation
domination era, during the reheating era and for a short period of
time.

\begin{figure}
\centering
\includegraphics[width=18pc]{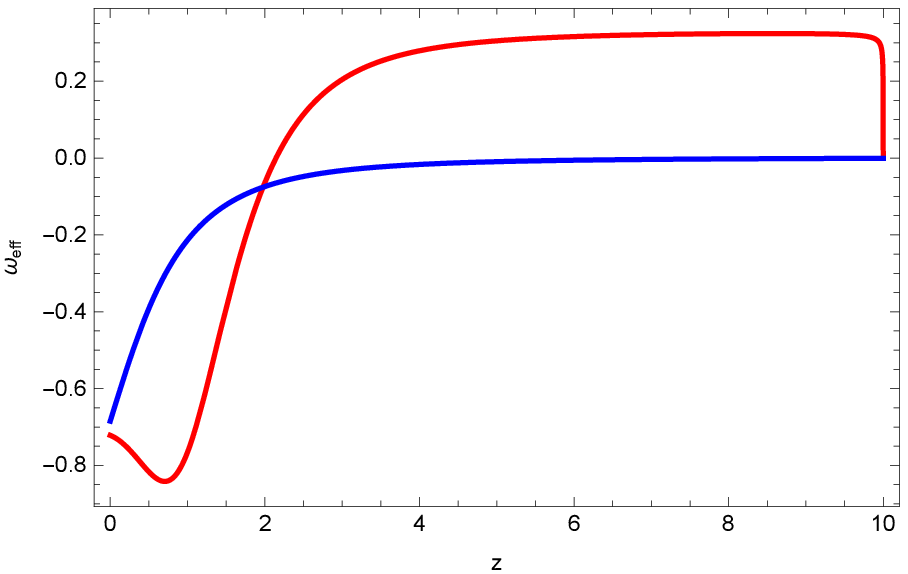}
\includegraphics[width=18pc]{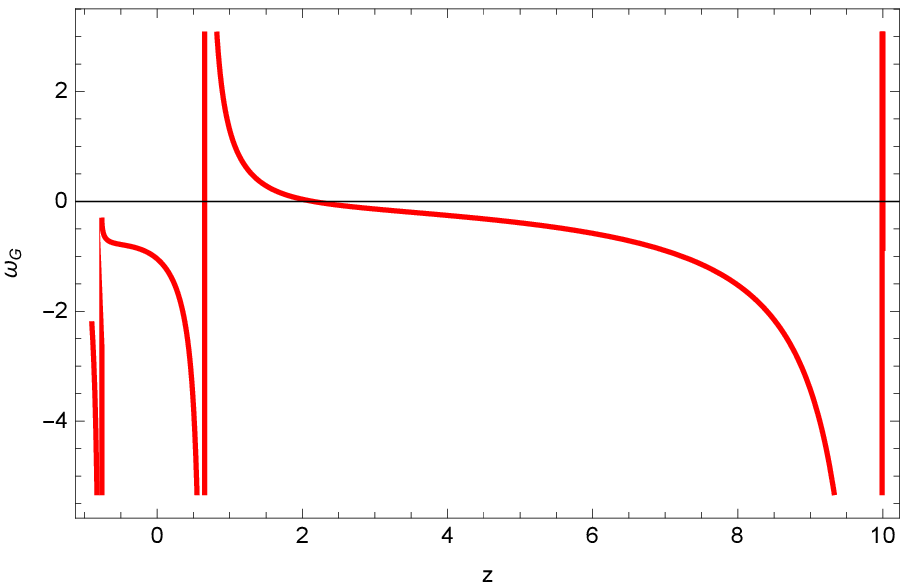}
\caption{The total EoS parameter $\omega_{tot}$ (left plot) and
the dark energy EoS parameter $\omega_{DE}(z)$ (right plot) as
functions of the redshift. The $f(R)$ model corresponds to red
curves and the $\Lambda$CDM model to blue curves.}\label{plot2}
\end{figure}

\begin{figure}
\centering
\includegraphics[width=18pc]{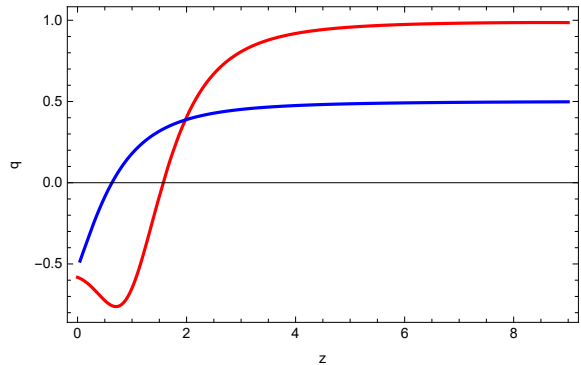}
\caption{The deceleration parameter, versus the redshift for the
deformed f(R) model (red curve) and for the $\Lambda$CDM model
(blue curve).} \label{plot3}
\end{figure}

Now let us consider the late-time evolution of the model, which is
affected by the dark energy $f(R)$ gravity term and by the
radiation and matter perfect fluids. Considering the dark energy
$f(R)$ gravity term, the parameters $\Lambda_1$ and $\Lambda_2$
will be considered to be of the order of the present day
cosmological constant. For the numerical analysis that will
follow, we specifically assume that $\Lambda_1\sim 19.5\Lambda$
and $\Lambda_2\sim 15 \Lambda$, where $\Lambda\simeq 11.895\times
10^{-67}$eV$^2$. Also the parameter $\delta$ in Eq. (\ref{frde})
will be chosen to be $\delta=2.01$. With this fine tuning, the
late-time era will prove to be viable and compatible with several
observational constraints on the dark energy era. Now let us
quantify the late-time era description of our model, firstly
considering the field equations in the presence of matter fluids,
which can be written in the Einstein-Hilbert form as follows,
\begin{align}\label{flat}
& 3H^2=\kappa^2\rho_{tot}\, ,\\ \notag &
-2\dot{H}=\kappa^2(\rho_{tot}+P_{tot})\, ,
\end{align}
where in the case at hand, the total energy density is equal to
$\rho_{tot}=\rho_{m}+\rho_r$. The energy density of dark energy
$\rho_{DE}$ is a geometric energy density contribution, cause by
$f(R)$ gravity and controls the late-time dynamics. The dark
energy density is defined to be,
\begin{equation}\label{degeometricfluid}
\kappa^2\rho_{DE}=\frac{f_R R-f}{2}+3H^2(1-f_R)-3H\dot{f}_R\, ,
\end{equation}
and the corresponding  dark energy pressure is defined as,
\begin{equation}\label{pressuregeometry}
\kappa^2P_{DE}=\ddot{f}_R-H\dot{f}_R+2\dot{H}(f_R-1)-\kappa^2\rho_{DE}\,
,
\end{equation}
while the total pressure is $P_{tot}=P_r+P_{DE}$. We shall rewrite
the field equations in terms of the redshift parameter
$1+z=\frac{1}{a}$ and the statefinder function $y_H(z)$
\cite{Bamba:2012qi,Odintsov:2020vjb,Odintsov:2020qyw,reviews1},
\begin{equation}\label{yHdefinition}
y_H(z)=\frac{\rho_{DE}}{\rho^{(0)}_m}\, ,
\end{equation}
where $\rho^{(0)}_m$ stands for the energy density of cold dark
matter today. The function $y_H(z)$ is written as follows,
\begin{equation}\label{finalexpressionyHz}
y_H(z)=\frac{H^2}{m_s^2}-(1+z)^{3}-\chi (1+z)^4\, ,
\end{equation}
where $\rho_m=\rho^{(0)}_m (1+z)^3$ and also $\chi$ is defined as
$\chi=\frac{\rho^{(0)}_r}{\rho^{(0)}_m}\simeq 3.1\times 10^{-4}$,
with $\rho^{(0)}_r$ being the radiation energy density today. We
also introduced the parameter $m_s$ defined as
$m_s^2=\frac{\kappa^2\rho^{(0)}_m}{3}=H_0\Omega_c=1.37201\times
10^{-67}$eV$^2$, and the Hubble rate is constrained by the CMB to
have approximately the value $H_0\simeq 1.37187\times 10^{-33}$eV
\cite{Planck:2018vyg}. The Friedmann equation in terms of the
redshift is written as,
\begin{equation}\label{differentialequationmain}
\frac{d^2y_H(z)}{d z^2}+J_1\frac{d y_H(z)}{d z}+J_2y_H(z)+J_3=0\,
,
\end{equation}
where the functions $J_1$, $J_2$ and $J_3$ are defined as follows,
\begin{align}\label{diffequation}
& J_1=\frac{1}{z+1}\left(
-3-\frac{1-F_R}{\left(y_H(z)+(z+1)^3+\chi (1+z)^4\right) 6
m_s^2F_{RR}} \right)\, , \\ \notag & J_2=\frac{1}{(z+1)^2}\left(
\frac{2-F_R}{\left(y_H(z)+(z+1)^3+\chi (1+z)^4\right) 3
m_s^2F_{RR}} \right)\, ,\\ \notag & J_3=-3(z+1)-\frac{\left(1-F_R
\right)\Big{(}(z+1)^3+2\chi (1+z)^4
\Big{)}+\frac{R-F}{3m_s^2}}{(1+z)^2\Big{(}y_H(z)+(1+z)^3+\chi(1+z)^4\Big{)}6m_s^2F_{RR}}\,
,
\end{align}
and also we defined $f(R)=R+F(R)$ and $F_{RR}=\frac{\partial^2
F}{\partial R^2}$, while $F_{R}=\frac{\partial F}{\partial R}$. In
our case, $F(R)=f_{DE}(R)$, with $f_{DE}(R)$ being defined in Eq.
(\ref{frde}). For our numerical analysis of the late-time era, the
following initial conditions shall be assumed in the redshift
interval $z=[0,10]$,
\begin{equation}\label{generalinitialconditions}
y_H(z_f)=\frac{\Lambda}{3m_s^2}\left(
1+\frac{(1+z_f)}{1000}\right)\, , \,\,\,\frac{d y_H(z)}{d
z}\Big{|}_{z=z_f}=\frac{1}{1000}\frac{\Lambda}{3m_s^2}\, ,
\end{equation}
which are physically motivated by matter domination era
\cite{Bamba:2012qi,Odintsov:2020vjb,Odintsov:2020qyw,reviews1}. In
terms of the statefinder function $y_H(z)$, the Ricci scalar is
written as,
\begin{equation}\label{ricciscalarasfunctionofz}
R(z)=3m_s^2\left( 4y_H(z)-(z+1)\frac{d y_H(z)}{d
z}+(z+1)^3\right)\, .
\end{equation}
and accordingly the dark energy density parameter $\Omega_{DE}$
is,
\begin{equation}\label{omegaglarge}
\Omega_{DE}(z)=\frac{y_H(z)}{y_H(z)+(z+1)^3+\chi (z+1)^4}\, .
\end{equation}
Also the dark energy EoS parameter $\omega_{DE}$, is written in
terms of $y_H(z)$ as follows,
\begin{equation}\label{omegade}
\omega_{DE}(z)=-1+\frac{1}{3}(z+1)\frac{1}{y_H(z)}\frac{d
y_H(z)}{d z}\, ,
\end{equation}
and the total EoS parameter takes the form,
\begin{equation}\label{totaleosparameter}
\omega_{tot}(z)=\frac{2 (z+1) H'(z)}{3 H(z)}-1\, .
\end{equation}
The dark energy EoS parameter and the dark energy density
parameter are constrained by the Planck 2018 constraints
\cite{Planck:2018vyg}, so these are important quantities. Finally,
we shall consider the deceleration parameter $q$ in order to
compare our model with the $\Lambda$CDM model, which is defined as
follows,
\begin{align}\label{statefinders}
& q=-1-\frac{\dot{H}}{H^2}=-1+(z+1)\frac{H'(z)}{H(z)}\, .
\end{align}
Finally, the results of our numerical analysis shall be compared
with the base $\Lambda$CDM model for which the Hubble rate is,
\begin{equation}\label{lambdacdmhubblerate}
H_{\Lambda}(z)=H_0\sqrt{\Omega_{\Lambda}+\Omega_M(z+1)^3+\Omega_r(1+z)^4}\,
,
\end{equation}
where $\Omega_{\Lambda}\simeq 0.681369$ and $\Omega_M\sim 0.3153$
\cite{Planck:2018vyg}, while $\Omega_r/\Omega_M\simeq \chi$. By
numerically solving the differential equation
(\ref{differentialequationmain}), we obtain the function $y_H(z)$
and from it the Hubble rate $H(z)$. The results of our analysis
are presented in Figs. \ref{plot2} and \ref{plot3}. In Fig.
\ref{plot2} we plot the total EoS parameter $\omega_{tot}$ (left
plot) and the dark energy EoS parameter (right plot) as functions
of the redshift. In Fig. \ref{plot3} we plot the deceleration
parameter versus the redshift and in all the plots, the red curves
correspond to the $f(R)$ model, while the blue curves to the base
$\Lambda$CDM model. Overall, the $f(R)$ gravity model
qualitatively behaves as the $\Lambda$CDM model, but it is surely
distinct from it. At present day however, the $f(R)$ model is
quite similar with the $\Lambda$CDM model.
Also quite good compatibility properties with the Planck
constraints \cite{Planck:2018vyg} are obtained for the $f(R)$
gravity model, since for the $f(R)$ model yields
$\omega_{DE}(0)\simeq -1.04573$ and the Planck constraints are
$\omega_{DE}=-1.018\pm 0.031$. Also with regard to the dark energy
density parameter the $f(R)$ model yields $\Omega_{DE}(0)\simeq
0.690065$ and the Planck constraints are $\Omega_{DE}=0.6847\pm
0.0073$.
The deceleration parameter for the $f(R)$ model at present day is
$q(0)=-0.582385$, while for the $\Lambda$CDM model is
$q(0)=-0.52701$. Hence the $f(R)$ gravity model at hand behaves
quite similarly to the $\Lambda$CDM and is deemed a viable dark
energy model. Having described the qualitative behavior of the
Universe in the various patches of its evolution, in the next
section we shall consider the produced energy spectrum of the
primordial gravitational waves. Specifically we shall show that
the geometrically originating short in duration abnormal reheating
era can significantly enhance the energy spectrum of the
primordial gravitational waves. This is to be contrasted with the
GR case, where a similar abnormal reheating era does not amplify
the energy spectrum of the gravitational waves significantly.

\section{Primordial Gravitational Wave Energy Spectrum Amplification Due to non-canonical $f(R)$ Reheating}

\begin{figure}[h!]
\centering
\includegraphics[width=40pc]{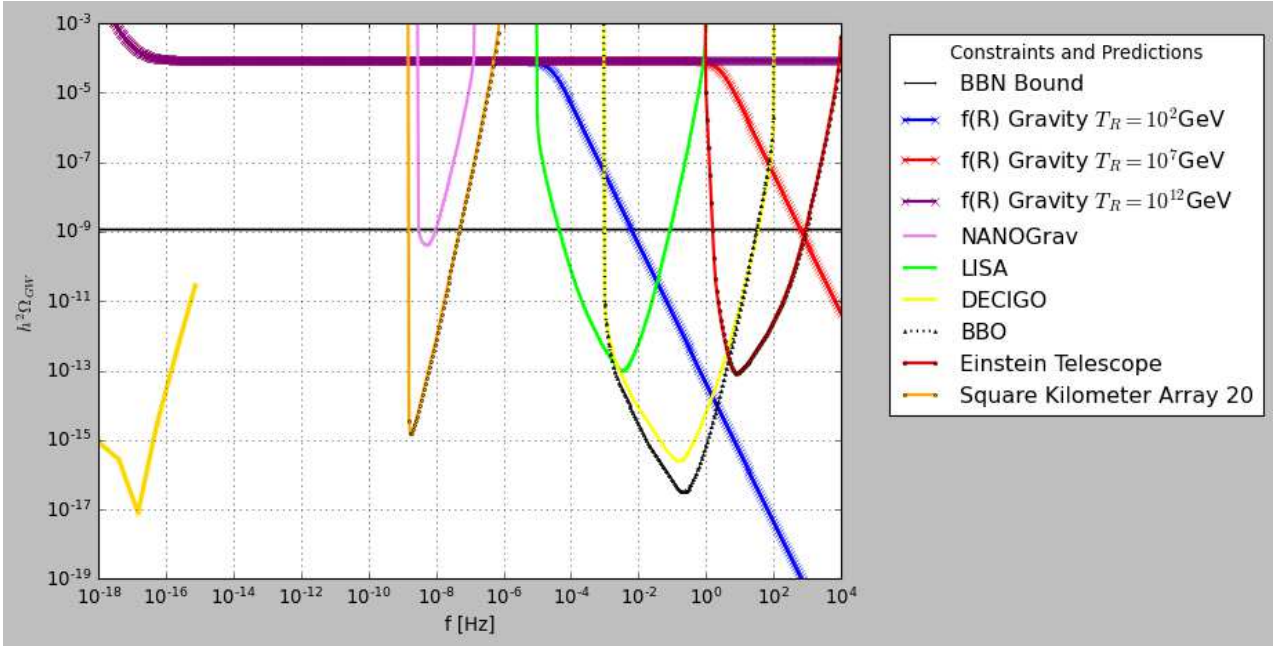}
\caption{The $h^2$-scaled gravitational wave energy spectrum for
pure $f(R)$ gravity. The $f(R)$ gravity curves correspond to three
different reheating temperatures, the purple curve to
$T_R=10^{12}$GeV, the red curve to $T_R=10^{7}$GeV and the blue
curve to $T_R=10^{2}$GeV. The $f(R)$ gravity gravitational wave
energy spectrum is significantly amplified due to the
geometrically originating short abnormal reheating era.}
\label{plotfinalfrpure}
\end{figure}

As we mentioned in the introduction, in about a decade from now,
several experiments will probe directly the inflationary era,
seeking for a stochastic background of primordial gravitational
waves. Thus, the theoretical predictions on the energy spectrum of
primordial gravitational waves which have been developing for
decades
\cite{Boyle:2005se,Denissenya:2018mqs,Turner:1993vb,Schutz:2010xm,Sathyaprakash:2009xs,Caprini:2018mtu,
Kuroyanagi:2008ye,Clarke:2020bil,Nakayama:2009ce,Smith:2005mm,Giovannini:2008tm,
Liu:2015psa,Zhao:2013bba,Vagnozzi:2020gtf,Watanabe:2006qe,Kamionkowski:1993fg,Giare:2020vss,
Nishizawa:2017nef,Arai:2017hxj,Nunes:2018zot,Campeti:2020xwn,
Zhao:2006eb,Cheng:2021nyo,Chongchitnan:2006pe,Lasky:2015lej,Guzzetti:2016mkm,Capozziello:2017vdi,Odintsov:2021kup,Benetti:2021uea,Cai:2021uup,Lin:2021vwc,Zhang:2021vak,Odintsov:2021urx},
will be put to test. The predicted energy spectrum of single
scalar field inflationary models and of $f(R)$ gravity is too
small to be detected from future experiments, thus in the case
that some future stochastic primordial signal is detected, these
two theories will not provide a good fit to the data. This
consideration however is based on the assumption that a standard
reheating and the subsequent radiation domination era take place.
In this scenario, the EoS parameter during the whole radiation
domination era is $w=1/3$. If a non-standard reheating era takes
place, the spectrum will be affected, see for example
\cite{Boyle:2005se}. Specifically, the EoS parameter during
reheating can take values $0<w<1/3$ \cite{Boyle:2005se} and an EoS
$w=0$ during reheating would mimic the effects of massive relics
on primordial gravitational waves \cite{Boyle:2005se}. In the
context of GR, the impact of a different from $w=1/3$ EoS
parameter during radiation domination, is small in magnitude
though, as we evince shortly. This is not true in the case that
the abnormal reheating era has a geometric origin, as we show in
this section. Indeed, we will show that if the abnormal reheating
is generated the $f(R)$ gravity (\ref{newsolutionsnoneulerssss})
of the previous section, the amplification of the primordial
gravitational wave energy spectrum is significant and can be
measurable by most of the future interferometers. Before we start
our analysis, we shall need to specify the duration of the
abnormal reheating era. As we assumed in Eq. (\ref{fralltimes})
the abnormal reheating era commences just after the end of the
inflationary era with temperature $T_e\sim 10^{12}$GeV and lasts
for a short period until the temperature drops to $T_a\sim
10^{10}$GeV. Now we should relate the temperatures to redshifts,
so using the approximate relation $T=T_0(1+z)$
\cite{Garcia-Bellido:1999qrp}, where $T_0$ is the present day
temperature $T_0=2.58651\times 10^{-4}$eV, we may obtain the
redshifts corresponding to the beginning and the end of the
abnormal reheating era. The temperature at the end of inflation
corresponds to $z_e=3.86621\times 10^{24}$ and the redshift at the
end of the abnormal reheating is $z_a=3.86621\times 10^{22}$.
After that, we assume that the standard radiation domination takes
place, where the radiation perfect fluid drives the evolution, and
subsequently the matter fluid and the $f_{DE}(R)$ drive the
late-time era. Let us now recall the formalism of $f(R)$ gravity
cosmological gravitational waves, see \cite{Odintsov:2021kup} and
references therein for details. For a perturbed flat FRW
background, the Fourier transformed tensor perturbation satisfies
the following ,
\begin{equation}\label{fouriertransformationoftensorperturbation}
\frac{1}{a^3f_R}\frac{{\rm} d}{{\rm d} t}\left(a^3f_R\dot{h}(k)
\right)+\frac{k^2}{a^2}h(k)=0\, ,
\end{equation}
or equivalently,
\begin{equation}\label{mainevolutiondiffeqnfrgravity}
\ddot{h}(k)+\left(3+\alpha_M
\right)H\dot{h}(k)+\frac{k^2}{a^2}h(k)=0\, ,
\end{equation}
where parameter $\alpha_M$ for a general $f(R)$ gravity is defined
as follows,
\begin{equation}\label{amfrgravity}
a_M=\frac{f_{RR}\dot{R}}{f_RH}\, .
\end{equation}
In order to quantify the $f(R)$ gravity effects on the
gravitational waves, we adopt a WKB approach
\cite{Nishizawa:2017nef,Arai:2017hxj}, in which the WKB solution
of the tensor perturbation differential equation reads,
\begin{equation}\label{mainsolutionwkb}
h=e^{-\mathcal{D}}h_{GR}\, ,
\end{equation}
where $h_{GR}$ is the GR waveform which corresponds to $a_M=0$.
The physical quantity $\mathcal{D}$ contains the $f(R)$ gravity
effects on the gravitational waves, and it is defined as,
\begin{equation}\label{dform}
\mathcal{D}=\frac{1}{2}\int^{\tau}a_M\mathcal{H}{\rm
d}\tau_1=\frac{1}{2}\int_0^z\frac{a_M}{1+z'}{\rm d z'}\, .
\end{equation}
In order to find the overall effect of $f(R)$ gravity on the
gravitational waves at present day, the quantity $\mathcal{D}$ has
to be evaluated from present day at $z=0$ until the end of
inflation at $z_e$. However, as we now show, the most important
integration periods are: from present day until recombination and
from the end of the abnormal reheating era, until its start. We
discuss this important issue shortly. The GR energy spectrum of
primordial gravitational waves is,
\begin{equation}
    \Omega_{\rm gw}(f)= \frac{k^2}{12H_0^2}\Delta_h^2(k),
    \label{GWspec}
\end{equation}
where $\Delta_h^2(k)$ is
\cite{Boyle:2005se,Nishizawa:2017nef,Arai:2017hxj,Nunes:2018zot,Liu:2015psa,Zhao:2013bba,Odintsov:2021kup},
\begin{equation}\label{mainfunctionforgravityenergyspectrum}
    \Delta_h^2(k)=\Delta_h^{({\rm p})}(k)^{2}
    \left ( \frac{\Omega_m}{\Omega_\Lambda} \right )^2
    \left ( \frac{g_*(T_{\rm in})}{g_{*0}} \right )
    \left ( \frac{g_{*s0}}{g_{*s}(T_{\rm in})} \right )^{4/3} \nonumber  \left (\overline{ \frac{3j_1(k\tau_0)}{k\tau_0} } \right )^2
    T_1^2\left ( x_{\rm eq} \right )
    T_2^2\left ( x_R \right ),
\end{equation}
where $\Delta_h^{({\rm p})}(k)^{2}$ stands for the inflationary
tensor power spectrum,
\begin{equation}\label{primordialtensorpowerspectrum}
\Delta_h^{({\rm
p})}(k)^{2}=\mathcal{A}_T(k_{ref})\left(\frac{k}{k_{ref}}
\right)^{n_T}\, ,
\end{equation}
evaluated at the CMB pivot scale $k_{ref}=0.002$$\,$Mpc$^{-1}$.
Also, $n_T$ denotes the inflationary tensor spectral index and
$\mathcal{A}_T(k_{ref})$ stands for amplitude of the tensor
perturbations, which is,
\begin{equation}\label{amplitudeoftensorperturbations}
\mathcal{A}_T(k_{ref})=r\mathcal{P}_{\zeta}(k_{ref})\, ,
\end{equation}
where $r$ denotes the tensor-to-scalar ratio and
$\mathcal{P}_{\zeta}(k_{ref})$ denotes the amplitude the scalar
perturbations. Thus finally, we have,
\begin{equation}\label{primordialtensorspectrum}
\Delta_h^{({\rm
p})}(k)^{2}=r\mathcal{P}_{\zeta}(k_{ref})\left(\frac{k}{k_{ref}}
\right)^{n_T}\, .,
\end{equation}
hence the energy spectrum of the primordial gravitational waves
for the GR and the $f(R)$ gravity waveforms takes the form,
\begin{align}
\label{GWspecfR}
    &\Omega_{\rm gw}(f)=e^{-2\mathcal{D}}\times \frac{k^2}{12H_0^2}r\mathcal{P}_{\zeta}(k_{ref})\left(\frac{k}{k_{ref}}
\right)^{n_T} \left ( \frac{\Omega_m}{\Omega_\Lambda} \right )^2
    \left ( \frac{g_*(T_{\rm in})}{g_{*0}} \right )
    \left ( \frac{g_{*s0}}{g_{*s}(T_{\rm in})} \right )^{4/3} \nonumber  \left (\overline{ \frac{3j_1(k\tau_0)}{k\tau_0} } \right )^2
    T_1^2\left ( x_{\rm eq} \right )
    T_2^2\left ( x_R \right )\, ,
\end{align}
where $\mathcal{D}$ can be found in Eq. (\ref{dform}) and let us
now discuss the redshift intervals for which it will be evaluated.
Apparently, the redshift intervals will be from present day up to
a redshift where the Universe entered the radiation domination
era, and from $z_a$ when the abnormal reheating era ends, up to
$z_e$ when inflation ends and the abnormal reheating era starts.
For the two intervals, the dominant $f(R)$ gravity which drives
the evolution is different, for example for the first interval
$R+f_{DE}(R)$ drives the evolution, while for the second interval,
$f_p(R)$ appearing in Eq. (\ref{newsolutionsnoneulerssss}) drives
the evolution. In both cases, the $f(R)$ gravity synergistically
with matter and radiation fluids drive the evolution of the
Universe. Thus the parameter $\mathcal{D}$ has to be numerically
evaluated in the following way,
\begin{equation}\label{dformexplicitcalculation}
\mathcal{D}=\frac{1}{2}\left(\int_0^{z_a}\frac{a_{M_1}}{1+z'}{\rm
d z'}+\int_{z_a}^{z_e}\frac{a_{M_2}}{1+z'}{\rm d z'}\right)\, ,
\end{equation}
with the parameter $a_{M_1}$ and $a_{M_2}$ being defined in Eq.
(\ref{amfrgravity}), but with $a_{M_1}$ evaluated for
$f(R)=R+f_{DE}(R)$ and $a_{M_2}$ being evaluated for $f_p(R)$
appearing in Eq. (\ref{newsolutionsnoneulerssss}). For the first
integration, the integral $\int_0^{z_a}\frac{a_{M_1}}{1+z'}{\rm d
z'}$ receives contribution from present day up to the
recombination era but it is truly minor and of the order $\sim
10^{-8}$. Beyond the recombination redshift the numerical
integration yields zero, and by also considering that when the
Universe enters the radiation domination era, $\dot{R}=0$, this
means that the integral $\int_0^{z_a}\frac{a_{M_1}}{1+z'}{\rm d
z'}$ vanishes up to redshift $z_a$. This feature is model
dependent, however this is also true in most of the dark energy
models of $f(R)$ gravity, see the discussion and models of
\cite{Odintsov:2021kup}. Thus the only significant contribution to
the parameter $\mathcal{D}$ comes from the second integral, namely
$\int_{z_a}^{z_e}\frac{a_{M_2}}{1+z'}{\rm d z'}$ which is
evaluated to be $\int_{z_a}^{z_e}\frac{a_{M_2}}{1+z'}{\rm d
z'}\sim -33$, thus an amplification of the gravitational wave
energy spectrum occurs of the order $\mathcal{O}(10^{14})$ occurs.
Hence, overall the signal is significantly amplified and thus
detectable by all the future experiments seeking high frequency
gravitational waves. This can be seen in Fig.
\ref{plotfinalfrpure} where we present the predicted $f(R)$
gravity $h^2$-scaled energy spectrum with purple, blue and red
curves and also the sensitivity curves of most of the future high
frequency experiments, and also the low frequency Litebird
experiment. The $f(R)$ gravity curves correspond to three
different reheating temperatures, and specifically, the purple
curve to $T_R=10^{12}$GeV, the red curve to $T_R=10^{7}$GeV and
the blue curve to $T_R=10^{2}$GeV. The three curves are
indistinguishable up to frequencies of the order
$\mathcal{O}(10^{-4})$Hz and thereafter the low reheating
temperature blue curve breaks off, while the red curve breaks off
around $\mathcal{O}(1)$Hz.
Thus the effect of a geometrically originating short lasting
abnormal reheating era, with $w=0.1$ results to a significant
amplification of the gravitational wave energy spectrum. We need
to note two things: firstly the amplification occurs for other
values of $w$ too, but the final amplification is different of
course. Secondly the duration of this abnormal reheating era
greatly affects the amplification of the gravitational wave energy
spectrum, and the longer the abnormal reheating era lasts, the
larger the amplification is. Also at this point, let us note that
the $f(R)$ gravity effect in the whole process is significant, and
this is in contrast with the GR situation. Indeed, in the latter
case, the change in the gravitational wave energy spectrum in the
case that the background EoS parameter is $w\neq 1/3$ is a
multiplicative factor $\sim \frac{\Gamma
(\frac{2}{1+3w})}{\pi}\left(1+3w\right)^{\frac{4}{1+3w}}$
\cite{Boyle:2005se}. Thus for $w=0.1$ one gets an overall damping
of the order $\mathcal{O}(1/2)$ in the gravitational wave energy
spectrum. Hence the effect of a geometric term like $f(R)$
gravity, which causes a deviation from the standard $w=1/3$
radiation domination era pattern, can be quite significant and
measurable. Thus if a signal is detected in some future high
frequency experiment, the scenario of an abnormal reheating era
caused by some higher curvature gravity might play a prominent
role for the identification of the theory that caused signal. Even
in this case though, the road toward understanding the exact
theory behind a future signal, is long and thorny.

Before closing this section, we need to address another important
issue related to the superhorizon modes of the modified gravity we
considered in this paper. So far we considered the subhorizon
modes and the effects caused on these modes by an abnormal
reheating era with EoS parameter $w$. We need to clarify an
important issue before discussing any possible effects of the
abnormal reheating era on the superhorizon modes. The subhorizon
modes are the ones with small wavelength which were subhorizon
modes during inflation, basically they became subhorizon just
after inflation started. Eventually these were the first modes
that reentered the horizon after inflation ended. Exactly for
these modes, the WKB method we used applies, so we were able to
measure the overall amplification caused by the abnormal reheating
era on them. On the antipode of these modes, lie the superhorizon
modes, and specifically the ones with wavelength $\lambda \geq
10\,$Mpc, correspond to modes probed by the CMB experiments and
the LiteBird experiment. These modes were superhorizon during
reheating and during the inflationary era, and became subhorizon
during the era probed by the CMB experiments and the LiteBird
experiment. Now interestingly enough, one may ask the reasonable
question, is the enhancement of the superhorizon modes comparable
to the one obtained for the subhorizon modes? In such a case, one
may claim that scenarios which will be probed by the
high-frequency experiments are already ruled out by the CMB
experiments, since the LiteBird curve in the sensitivity curves
lies below the direct detection curves of the high-frequency
gravitational wave experiments.  Also notably, similar
considerations apply for previous CMB experiments as well, see for
example Fig. 2 in Ref. \cite{Caldwell:2019vru}.

This question is very interesting so let us address in brief the
superhorizon modes evolution, for the whole time that these are
actually superhorizon modes. As we already mentioned, the
superhorizon modes relevant to the CMB experiments are those with
$\lambda \geq 10\,$Mpc, and these are superhorizon during the
whole inflationary era and remain superhorizon until $z\sim 1100$
which the redshift corresponding to the CMB. Thus the superhorizon
modes during the abnormal reheating era, are still superhorizon
modes. For these modes, since for superhorizon modes we have
$k\ll H a$, the evolution differential equation for the tensor
modes, namely Eq. (\ref{mainevolutiondiffeqnfrgravity}), becomes,
\begin{equation}\label{mainevolutiondiffeqnfrgravityup}
\ddot{h}_{\ell}(k)+\left(3+\alpha_M \right)H\dot{h}_{\ell}(k)=0\,
.
\end{equation}
The general solution of the above differential equation is,
\begin{equation}\label{generalsolutiondiffeqn}
h_{\ell}(k)=C_{\ell}(k)+D_{\ell}(k)\int_1^t \exp
\left(\int_1^{\eta} (-a_M(\tau)-3 H(\tau)) \,
\mathrm{d}\tau\right) \, \mathrm{d}\eta \, ,
\end{equation}
thus it is apparent that the solution describes a time-independent
frozen term $C_{\ell}(k)$ and the second term in Eq.
(\ref{generalsolutiondiffeqn}), which is an exponentially decaying
mode. Hence during the whole abnormal reheating era, the
superhorizon modes remain frozen and thus there should be no
effect by the abnormal reheating era on them. However, at this
point it is crucial to note two things: first, if the contribution
from $a_M(\tau)$ during the reheating era is negative in Eq.
(\ref{generalsolutiondiffeqn}), then it is possible that
superhorizon do not freeze and evolve. This would break the linear
approximation though, and such effects should be carefully
addressed. In this paper though we did not consider these modes,
but we aimed for the study of the subhorizon modes, so this study
is deferred to a future work. Secondly, it should be noted that
for these low-frequency-large wavelength modes, several effects
caused by free-streaming relativistic particles like neutrinos
should be taken into account, which however we did not take into
account. Certainly such effects should be taken into account, so
the part of the curves we presented in the plots we presented in
this section, should be corrected at low-frequencies.

\section{Conclusions}

In this paper we studied the effects of an short abnormal
reheating era generated by higher order curvature terms, on the
primordial gravitational wave energy spectrum. Specifically we
focused on $f(R)$ gravity theory and we discussed how $f(R)$
gravity may affect the various evolutionary patches of the
Universe. Specifically, $f(R)$ gravity may drive inflation,
ignoring perfect matter fluids, but can also affect the late dark
matter and the dark energy era, providing a $\Lambda$CDM-like
evolution in the presence of matter and radiation fluids. The era
for which $f(R)$ gravity may not play a significant role at all is
the radiation domination era for a flat FRW Universe. This however
only holds true in the case that the background EoS parameter is
that of radiation $w=1/3$, since for a FRW Universe, $R=0$ in this
case. If an alternative EoS parameter governs the early stages of
the radiation domination era, then $f(R)$ gravity might be the
actual generator of this era. We examined how and which $f(R)$
gravity can generate each evolutionary patch of the Universe, from
inflation, the abnormal reheating era, and finally the late-time
era. Actually the late-time era $f(R)$ gravity term can be present
from primordial times and has no significant effect on the
evolution, until late times where it drives the late-time era.
Using a WKB approach, valid for subhorizon modes, which are the
modes that will actually be probed by most high frequency future
experiments, we quantified the effect of the $f(R)$ gravity on the
primordial gravitational wave energy spectrum. We calculated
numerically the amplification factor, and the only significant
contribution came only from the short abnormal reheating era. In
fact, the longer this short abnormal reheating era lasts, the
larger the amplification of the signal is. Now let us briefly
discuss some possible future scenarios. The main problem of
non-tachyon single scalar field theories, and of their Jordan
frame counterpart theories, namely $f(R)$ gravity, is that they
predict an undetectable by the future experiments signal. Thus if
inflation is verified in the CMB, and also a signal of a
stochastic gravitational wave background is detected by future
experiments, this would rule out both single scalar and $f(R)$
gravities. However, with the present paper we showed that in the
context of $f(R)$ gravity, alternative scenarios related to the
mysterious radiation domination era, might enhance significantly
the predicted signal. Similar scenarios were discussed briefly in
\cite{Boyle:2005se}, in the context of GR, however in GR the
amplification-damping is insignificant compared to the
geometrically originating on generated by $f(R)$ gravity. Hence
one may not rule out easily $f(R)$ gravity, and the quest for
future theorists and observational cosmologists, is to distinguish
the theory that produces a detected signal. Is it a blue-tilted
theory or it is a modified gravity theory with abnormal radiation
domination era? This is a difficult question to answer, and for
the moment model dependent, but the clue point will be from how
many interferometer experiments will the signal be captured. These
questions will possibly occupy the minds of theorists and
experimentalists for the next two decades.

\section*{Acknowledgments}

This work was supported by MINECO (Spain), project
PID2019-104397GB-I00 (S.D.O).

\end{document}